\documentclass[preprint,prd,amsmath,amssymb]{revtex4}
\usepackage{graphicx}
\usepackage{subfigure}
\usepackage{dcolumn}

\newcommand{\be}{\begin{equation}}
\newcommand{\ee}{\end{equation}}

\newcommand{\Tr}{{\rm Tr}}
\newcommand{\bea}{\begin{eqnarray}}
\newcommand{\eea}{\end{eqnarray}}
\newcommand{\bml}{\begin{multline}}
\newcommand{\eml}{\end{multline}}

\newcommand{\LE}{{\cal E}}

\newcommand{\LH}{{\cal H}}

\newcommand{\nn}{\nonumber}
\newcommand{\bm}[1]{\mbox{\boldmath $#1$}}
\newcommand{\bms}[1]{\mbox{\boldmath ${\scriptstyle #1}$}}
\newcommand{\hbm}[1]{\hat{\mbox{\boldmath $#1$}}}

\newcommand{\ord}{{\cal O}}

\newcommand{\link}{\begin{array}{l}\begin{picture}(85,125)
        \put ( 0, 0 )   {\circle{1}}
        \put ( 0,40 )   {\circle{1}}
        \put ( 0,80 )   {\circle{1}}

        \put (40, 0 )   {\circle{1}}
        \put (40,40 )   {\circle{1}}
        \put (40,80 )   {\circle{1}}

        \put (80, 0 )   {\circle{1}}
        \put (80,40 )   {\circle{1}}
        \put (80,80 )   {\circle{1}}

        \put (120, 0)   {\circle{1}}
        \put (120,40)   {\circle{1}}
        \put (120,80)   {\circle{1}}
        
        \put (36,28) {$x$}
        \put (70,28) {$x+a\mu$}

        \put (40,40) {\vector(1,0){23}}
        \put (40,40) {\line(1,0){40}}

\end{pictur}\end{array}}

\newcommand{\boldx}{\boldsymbol{x}}
\newcommand{\eqn}[1]{Eq.~(\ref{#1})}

\newcommand{\eqns}[2]{Eqs.~(\ref{#1}) and (\ref{#2})}

\newcommand{\fig}[1]{Fig.~\ref{#1}}

\newcommand{\sect}[1]{Section~\ref{#1}}

\newcommand{\rcite}[1]{Ref.~\onlinecite{#1}}
\newcommand{\rcites}[2]{Refs.~\onlinecite{#1} and~\onlinecite{#2}}

\begin{document}
\title{SU($N$) Lattice Gauge Theory on a Single Cube}

\author{Jesse Carlsson}
\email{j.carlsson@physics.unimelb.edu.au}
\author{Bruce H.~J.~McKellar}
\email{b.mckellar@physics.unimelb.edu.au}
\affiliation{School of Physics, The University of Melbourne}
\date{\today}
\begin{abstract}
In this paper we study the viability of persuing analytic variational
techniques for the calculation of glueball masses in 3+1 dimensional
Hamiltonian lattice gauge theory (LGT) in the pure gauge sector. 
We discuss the major problems
presented by a move from 2+1 to 3+1 dimensions and develop analytic
techniques to approximate the integrals appearing in 3+1 dimensional
variational glueball mass calculations. We calculate $0^{++}$ and
$1^{+-}$ glueball masses on a lattice consisting of a single cube. 
Despite the use of a very simplistic model, promising signs of an
approach to asymptotic scaling is displayed by the SU($N$) $1^{+-}$
glueball mass as $N$ is increased.  
\end{abstract}

\maketitle

\section{Outline}

In this paper we explore the viability of extending the analytic
techniques used with success in \rcites{thesis-paper1}{thesis-paper2}
to the calculation of glueball masses in the pure gauge sector in 3+1 
dimensions. The primary difficulty lies in the calculation of
expectation values in 3+1 dimensions. In \sect{3+1introduction} we
briefly review what has been achieved in Hamiltonian
LGT~\cite{Kogut:1975ag} in 3+1 dimensions. 
We discuss the difficulties faced in 3+1 dimensions and possible
solutions in \sect{bianchiidentities}. In
\sect{gausslaw} we consider the problem of Gauss' law
constraints. This is a topic that has been discussed in the context of
Hamiltonian LGT most recently by Ligterink, Walet and
Bishop~\cite{Ligterink:2000ug} and concerns the constraint equations
that appear when non-abelian gauge theories are canonically quantised. In
\sect{onecube} we move on to the calculation of variational
glueball masses on a single cube using the analytic variational technique
discussed in \rcite{thesis-paper1}. We finish in \sect{futurework} with a
discussion of the viability of pursuing analytic techniques for pure
SU($N$) LGT in 3+1 dimensions based on the results of \sect{onecube}.

\section{Introduction}
\label{3+1introduction}

From a renormalisation point of view the key difference between 2+1
and 3+1 dimensional gauge theory lies in the units of the coupling
constant. 2+1 dimensional
gauge theory has a coupling constant, $e^2$, with the dimensions of
mass and so the coupling constant explicitly sets a mass scale for
calculations on the lattice. In contrast
the 3+1 dimensional coupling constant is dimensionless. This makes the
extraction of continuum physics from lattice calculations more subtle
in 3+1 dimensions than in 2+1.

On a practical level, there is a more serious problem faced in moving
from 2+1 to 3+1 dimensions for Hamiltonian LGT calculations. 
The analytic techniques that were used
with success in \rcites{thesis-paper1}{thesis-paper2} are no longer applicable.
These techniques rely heavily upon the fact that in 2+1 dimensions a change of variables
from links to plaquettes has unit Jacobian. The form for the
equivalent Jacobian in 3+1 dimensions is considerably more
complicated. The most
comprehensive study of the change of variables from links to
plaquettes is due to Batrouni~\cite{Batrouni:1982bg,Batrouni:1983ch}. 
We discuss this change of variables in more detail in \sect{bianchiidentities}.
%In this study it was
%found that for Abelian gauge theories the Jacobian could be expressed
%in a seperable form in terms of independent products of plaquettes around a cube, with one
%factor for each cube on the lattice. This allows integrals on a three
%dimensional lattice to be separated into integrals over cubes. 
%For the non-abelian case a similar
%reduction for a general cubic lattice could not be found. To express
%the Jacobian in a separable form required lattices of a
%particular shape. The largest volume candidate is a tower of elementary cubes. 

The extension of the techniques used in \rcite{thesis-paper1} to 3+1 dimensions is not
straightforward. There are a number of immediate problems. Firstly,
since plaquettes are not independent variables in 3+1 dimensions one
can not automatically work in the infinite volume limit. In a precise
study one would need to calculate identical quantities on different
sized lattices and extrapolate to the infinite volume limit.
Secondly, in the context of analytic calculations, even on small
lattices the integrals
involved in the calculation of basic matrix elements are considerably
more complicated than those encountered in 2+1 dimensions. Such matrix
elements could in
principle be carried out analytically on small lattices but since the number
of integration variables increases quickly with the volume of the
lattice a calculation on even a $5^3$ lattice would seem
exceedingly difficult. How quickly the infinite volume limit is
reached will therefore determine the worth of pursuing analytic Hamiltonian 
methods in 3+1 dimensions. Finally, there is the complication of Gauss' law which we discuss in \sect{gausslaw}.

The only Hamiltonian techniques to have been applied with any
success to the case of SU(3) gauge theory in 3+1 dimensions have been
strong coupling expansions, the $t$-expansion and exponential wave
function methods. Each of which we now summarise.

Strong coupling perturbative techniques were used in the early days of
LGT in an attempt to bridge the gap between the
strong and weak coupling limits. Strong coupling expansions of the
Callan-Symanzik $\beta $ function were calculated  to $\ord(g^{-24})$~\cite{Kogut:1979vg,Kogut:1980sg}
and showed signs of interpolating smoothly between the strong and weak
coupling limits. This suggested that the $\beta$ function was a
smooth function of the coupling with its only zero at $g=0$, providing
a strong argument at the time for the continuum limit of LGT 
confining quarks. Corresponding strong coupling expressions for
glueball masses did not share the same success. Despite strong
coupling calculations to $\ord(g^{-28})$~\cite{Hamer:1989qm} scaling
was not observed in $0^{++}$, $1^{+-}$ or $2^{++}$ glueball masses. It was
later realised that a roughening transition prevents a smooth
crossover from strong to weak coupling physics (see
\rcite{Montvay:1994cy-1} and references within). 

The $t$-expansion was introduced by Horn and
Weinstein~\cite{Horn:1984bq} as an analytic method suitable for the
study of LGT in the Hamiltonian formulation. It has been applied in
the calculation of glueball masses in 3+1 dimensions for 
SU(2)~\cite{Horn:1985ax} and
SU(3)~\cite{VanDenDoel:1986bw,vandenDoel:1987xk,Horn:1991fz} LGT in
the pure gauge sector. More recently it has been used in an attempt to
calculate the lowest hadron masses~\cite{Horn:1993wb}. For each case 
however asymptotic scaling of masses was not directly observed. 
Extrapolation techniques
such as Pad{\'e} approximants were required to probe the weak coupling
region. The extrapolated mass ratio results agreed with Monte
Carlo estimates of the time.  

The coupled cluster method and related exponential wave function
techniques have received by far the most attention in
Hamiltonian LGT in recent years. Essentially these techniques aim to
solve the Kogut-Susskind eigenvalue equation by making a suitable
ansatz for the wave function.  The coupled cluster method was
originally constructed with applications in nuclear physics in
mind~\cite{Coester:1958,Coester:1960} but has since found the majority
of its applications in molecular physics~\cite{Bishop:1987}. Its application
in the context of Hamiltonian LGT is described in
\rcites{Schutte:1997du}{McKellar:2000zk}. The truncated
eigenvalue method, developed by Guo, Chen and Li~\cite{Guo:1994vq},
is another exponential wave function technique to have 
found application in Hamiltonian LGT.  

A number of groups have made considerable progress in the application
of exponential wave function techniques to Hamiltonian LGT in 3+1 dimensions.
While most studies have explored gauge groups other than SU(3) in less
than three dimensions, studies of SU(3) glueballs in 3+1 dimensions 
have commenced. 
Results tangent to the expected scaling form, indicating an approach
to scaling, have been
obtained for SU(3) pure gauge theory in a coupled cluster calculation by
Leonard~\cite{ConradPhD}. However convergence with increasing orders
appears to be slow. Finite order truncation errors appear to be under
more control in the truncated eigenvalue method. The first calculations of
3+1 dimensional SU(3) glueball masses with this method~\cite{Hu:1997ys}
 gave a $1^{+-}$ to $0^{++}$ mass ratio that was consistent with the
Monte Carlo results of the time. The agreement was not as good for the
$0^{--}$ to $0^{++}$ mass ratio. A convincing
demonstration of asymptotic scaling has not yet been produced in a
calculation of glueball masses for SU(3) in 3+1 dimensions. 

More promising 3+1 dimensional results  
have been obtained for higher dimensional gauge groups. Ironically
these results have been obtained with much simpler methods than either
of the exponential wave function methods described above. 
Asymptotic scaling of the lowest $0^{++}$ glueball mass has been
demonstrated by Chin, Long and Robson in a variational calculation on a small volume ($6^3$
sites) lattice for SU(5)
and SU(6)~\cite{Chin:1986fe}. This calculation follows the
variational technique described in \rcite{thesis-paper1} but uses
Monte Carlo rather than analytic techniques to calculate the required
expectation values. It uses only plaquette states in the minimisation
basis rather than the large basis of rectangular states used in \rcites{thesis-paper1}{thesis-paper2}. Naively one would hope
that the same method could be explored on similar sized lattices for
larger $N$, with additional states in the minimisation basis, using
the analytic techniques of \rcite{thesis-paper1}. 
We take this as our motivation for the studies
presented in this paper. It will become clear however that attempting a
similar calculation to Chin, Long and Robson on a single cube using 
analytic techniques presents a significant challenge.

\section{The Move to 3+1 Dimensions}
\label{bianchiidentities}

In LGT one usually works in 2+1 dimensions to test
a technique with the intention of later extending it to the physically
relevant 3+1 dimensions. This may be
justified in the Lagrangian approach where the time coordinate 
is treated on the same
footing as the spatial coordinates and adding another dimension equates 
to nothing more
than an increased load on computer memory. However, in Hamiltonian LGT
two serious technical differences exists between 2+1 and 3+1
dimensions. The first is Gauss' law and the second is related to
constructing a Jacobian for transforming from link to plaquette variables.

To understand the first difference one needs to recall that Hamiltonian LGT
is formulated in the temporal gauge which sets $A_0 =0$. If one starts
with the Yang-Mills 
Lagrangian and performs the standard equal-time quantisation,
one runs into problems because the time derivative of $A_0$ does not
appear in the Lagrangian. The variational principle gives equations of
motion for the space-like components resulting in the standard 
Yang-Mills Hamiltonian. For the time-like component we obtain a set of
algebraic  
constraint equations which are the analogue of Gauss' law. There
is one constraint equation for each colour component of $A_0$, $N^2-1$ for
SU($N$), at each lattice site. As has been pointed out in the context of Hamiltonian LGT by Ligterink, Walet and 
Bishop~\cite{Ligterink:2000ug}, it is only when one
works with a set of variables whose
number of degrees of freedom matches the number of unconstrained
degrees of freedom in the theory, that one can avoid the
technicalities of constraint equations. For this reason, in
2+1 dimensions we really are quite lucky. The number of plaquette variables
on a square two dimensional lattice is  precisely equal to 
the number of unconstrained variables~\cite{Ligterink:2000ug}. 
For a three dimensional cubic lattice this is not the case. One could envisage
constructing a polyhedral lattice in $d$ dimensions such that the number of
faces (plaquettes) would equal the number of unconstrained variables. 
However such lattices appear to be prohibited
by Euler's equation which relates the number of vertices, edges and
faces of polyhedra.

The other serious technical difference between 2+1 and 3+1 dimensions, that of
constructing a Jacobian for transforming from link to plaquette
variables, is 
relevant to both the Hamiltonian and Lagrangian formulations of
LGT. It only becomes important when the method of choice relies on
plaquette variables for its implementation. The transformation from
link variables to plaquette variables is required in the approach taken
in \rcites{thesis-paper1}{thesis-paper2} to make use of the 
analytic results available for certain group integrals. Such a 
transformation does not
need to be made if one is happy to use Monte Carlo techniques to
handle the integrals as is the case in \rcite{Chin:1986fe}.

The most complete treatment of the
transformation from link to plaquette variables is due to
Batrouni~\cite{Batrouni:1983ch,Batrouni:1982bg} who worked in the 
Lagrangian formulation. His approach was based
on the continuum
work of Halpern~\cite{Halpern:1979ik} who constructed a field-strength 
formulation of
gauge theory. In Halpern's construction the definition of the field strength $F_{\mu \nu}$ is
inverted to give an expression for $A_\mu$ as a function of the field
strength. To do this requires the choice of a suitable gauge. The
Jacobian of the transformation is precisely the Bianchi identity; a
constraint equation on $F_{\mu \nu}$. For the Abelian case the
Bianchi identity is equivalent to the requirement that the total
magnetic flux leaving a volume is zero. Batrouni developed an equivalent
field strength formulation on the lattice. On the lattice, the link
elements, $U_l$, correspond to the vector potentials, $A_\mu$, and the 
plaquette variables correspond to the field strengths, $F_{\mu \nu}$. Batrouni
demonstrated that the Jacobian of the transformation from link
variables to plaquette variables was the lattice analogue of the
Bianchi identity. For Abelian gauge theories the lattice Bianchi
identity can be separated into factors which depend only on the plaquette variables of elementary cubes, with one factor for each cube of the lattice. 
For non-abelian
theories the Bianchi identity has only been found to separate in this
way for special types of lattices, the largest volume example being an infinite
tower of cubes. For the infinite lattice the Bianchi identity is a 
complicated nonlinear inseparable function of distant plaquette
variables. Interestingly it is the only source of correlations between
plaquette variables in LGT. Mean plaquette methods have been developed to 
deal with the added complications of the non-abelian Bianchi identity but 
have not progressed far~\cite{Batrouni:1982dx}.

To summarise, Hamiltonian LGT in 3+1 dimensions
faces some serious problems. Care must be taken in choosing
appropriate variables to work with if constraints on the lattice electric
fields are to be avoided. Additionally, if one wishes to use analytic 
techniques to calculate matrix elements the complications of the
Bianchi identity restricts the calculations to small lattices.

\section{Constraint Equations}
\label{gausslaw}

In this section we focus on one of the problems faced in moving 
from 2+1 to 3+1 dimensions, that of Gauss' law. From the
discussion of \sect{bianchiidentities} it would seem that one
needs either to construct an appropriate set of variables with
precisely the correct number of unconstrained degrees of freedom or be
faced with the problem of building constraint equations on the lattice 
electric fields. As has been pointed out by Ligterink, Walet and
Bishop~\cite{Ligterink:2000ug} there is an alternative. The problem of
satisfying Gauss' law can be solved by working with wave functions that
are annihilated by the generator of Gauss' law. We discuss this matter
in what follows.

The generator of Gauss' law on the lattice can be written as 
~\cite{Kogut:1975ag,Kogut:1980sg},
\bea
{\cal G}^a (\bm{x}) = \sum_{i} \left[\LE^a_i(\bm{x}) + \LE^a_{-i}(\bm{x})\right].
\eea   
Here $\LE^\alpha_i(\bm{x})$ is the lattice
chromoelectric field on the directed link running 
from $\bm{x}$ to $\bm{x}+a\hbm{i}$. 
In this notation the lattice electric fields satisfy the commutation relations~\cite{Kogut:1980sg}
\bea
\left[ \LE_i^a(\bm{x}) , U_j(\bm{y})\right] &=& T^a U_j(\bm{y}) \delta_{ij}
\delta_{\bms{x}\bms{y}} \label{commutations1}\\
\left[\LE_{-i}^a(\bm{x}+\hbm{i} a) , U_j(\bm{y})\right] &=& -
U_j(\bm{y})
T^a \delta_{ij} \delta_{\bms{x}\bms{y}} . \label{commutations2}
\eea 
Here $\hbm{i}$ is a unit vector in the $i$ direction and $\{T^a : 1\le a\le N^2-1\}$ is a basis for SU($N$). It is
common to use the Gell-Mann basis, in which case $T^a = \lambda^a/2$,
where $\{\lambda^a : 1\le a\le N^2-1\}$ is the set of traceless $N\times N$ Gell-Mann
matrices.
It should be pointed out that in this notation we have
$U^\dagger_i(\bm{x}) = U_{-i}(\bm{x}+\hbm{i} a)$. 
For physical states we must therefore have ${\cal G}^a (\bm{x}) |\psi\rangle =
0$ for each lattice site, $\bm{x}$, and all $a=1,\ldots,N^2-1$. It should be checked that this is
the case for the one plaquette exponential trial state,
\be
\begin{array}{c}\includegraphics{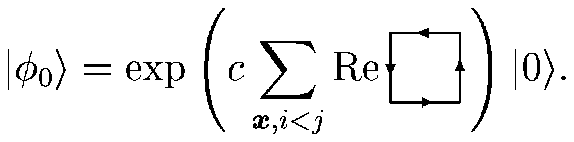}\end{array}
 \label{oneplaquette}
\ee
%\bea
%|\phi_0\rangle = \exp\left( c \sum_{\bms{x},i<j} \Real
% \plaquette \right)|0\rangle. \label{oneplaquette}
%\eea
Here, $|0\rangle $ is the strong coupling vacuum defined by
$\LE^\alpha_i(\bm{x})|0\rangle = 0$ for all $i$, $\bm{x}$ and $\alpha =
1,2,\ldots,N^2-1$. The directed square denotes the
traced ordered product of link 
operators, $U_i(\bm{x})$, around an elementary square, or plaquette, of the lattice,
\be
\begin{array}{c}\includegraphics{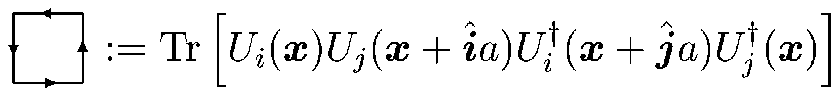}\end{array}
\ee
%\bea
%\plaquette := \Tr \left[ U_i(\bm{x})U_j(\bm{x}+\bm{i}a)
%U^\dagger_i(\bm{x}+\bm{j}a) U^\dagger_j(\bm{x})\right],
%\eea
where $a$ is the lattice spacing.

Consider first
a state consisting of a single plaquette acting on the strong coupling vacuum,
\bea
|p_{ij}(\bm{x})\rangle  = \Tr \left[ U_i(\bm{x}) U_j(\bm{x}+\hbm{i} a)
U_{-i}(\bm{x}+\hbm{j} a) U_{-j}(\bm{x})\right] |0\rangle.
\eea
Since $|0\rangle $ is annihilated by the electric field, using
\eqns{commutations1}{commutations2} we immediately have
\bea
{\cal G}^a(\bm{y})  |p_{ij}(\bm{x})\rangle = 0,
\eea
for all lattice sites, $\bm{y}$, not lying on the corners of the plaquette, 
$p_{ij}(\bm{x})$. Consider now sites lying on the corners of the plaquette in
question. In particular consider $\bm{y} = \bm{x}$. Making use of the
commutation relations of \eqns{commutations1}{commutations2} we have
\bea
{\cal G}^a(\bm{x})  |p_{ij}(\bm{x})\rangle &=& \Tr \left[ T^a U_i(\bm{x}) U_j(\bm{x}+\hbm{i} a)
U_{-i}(\bm{x}+\hbm{j} a) U_{-j}(\bm{x})\right] |0\rangle \nn\\
&&  - \Tr \left[ U_i(\bm{x}) U_j(\bm{x}+\hbm{i} a)
U_{-i}(\bm{x}+\hbm{j} a) U_{-j}(\bm{x}) T^a  \right] |0\rangle \nn\\
&=& 0.
\eea
The same applies for other sites on the plaquette. This argument
is not specific to plaquettes. Gauss' law is found to be satisfied
locally by any closed Wilson loop on the lattice, traced over colour
indices, acting on the strong coupling vacuum. It is easy to extend
this result  to products of such loops acting on $|0\rangle$. To see
this we consider how Gauss' law applies to the product of two plaquettes
\bea
|p_{ij}(\bm{x})p_{ij}(\bm{x}+a\hbm{i})\rangle &=&
 p_{ij}(\bm{x})p_{ij}(\bm{x}+a\hbm{i}) |0\rangle.
\eea
Only at the sites $\bm{x}+a\hbm{i}$
and $\bm{x}+a(\hbm{i}+\hbm{j})$ does this case differ from the single plaquette example . Let us consider Gauss' law at
$\bm{x}_{+}=\bm{x}+a\hbm{i}$. Once again, using the fact that the
strong coupling vacuum is annihilated by the lattice chromoelectric field, and \eqns{commutations1}{commutations2} we have 
\bea
{\cal G}^a(\bm{x}_+) |p_{ij}(\bm{x})p_{ij}(\bm{x}_{+})\rangle
&=& \sum_k \left[ \LE^a_k(\bm{x}_+) +  \LE^a_{-k}(\bm{x}_+),
p_{ij}(\bm{x})p_{ij}(\bm{x}_+)\right]|0\rangle \nn\\
&=& \sum_k\left\{ 
\left[
\LE^a_k(\bm{x}_+)+\LE^a_{-k}(\bm{x}_+),p_{ij}(\bm{x})\right]p_{ij}(\bm{x}_+)|0\rangle
\right. \nn\\
&&\left. + 
p_{ij}(\bm{x})\left[
\LE^a_k(\bm{x}_+)+\LE^a_{-k}(\bm{x}_+),p_{ij}(\bm{x}_+)\right]|0\rangle\right\}
\nn\\
&=& 0.
\eea
This result is easily extended to arbitrary products of closed Wilson
loops, traced over colour indices, acting on $|0\rangle
$. Consequently, any function of such loops acting on $| 0\rangle $ also satisfies
Gauss' law provided it admits a Taylor series expansion. The one
plaquette trial state of \eqn{oneplaquette} thus obeys Gauss' law and is therefore suitable
for use in simulating physical states. 

\section{The One Cube Universe}
\label{onecube}

In this section we adapt the analytic techniques used with success in 
\rcites{thesis-paper1}{thesis-paper2} for use in 3+1 dimensions. 
As a starting point we consider the case of a
single cube. This will serve as a test case to assess the
feasibility of a larger volume study. Our aim is to
use a small basis of states to calculate 3+1 dimensional SU($N$) glueball masses variationally, and check if an approach to the correct scaling form is
observed. We do not expect to achieve scaling with such a small lattice.
However, an approach to scaling would warrant an extended study on larger
lattices. We start with a general description of the approach. 
The starting point is the choice of trial state. As
in 2+1 dimensions~\cite{thesis-paper1,thesis-paper2} we choose to work
with the one plaquette trial state of \eqn{oneplaquette}.

Many Hamiltonians are available for LGT calculations~\cite{Carlsson:2001wp}. In
\rcite{thesis-paper1} we used Kogut-Susskind~\cite{Kogut:1975ag},
improved and tadpole improved Hamiltonians to calculate glueball
masses for $N\le 5$. Here we employ the simplest of these, the
Kogut-Susskind Hamiltonian, which is defined for pure SU($N$) gauge
theory with coupling, $g^2$, on a lattice with
spacing, $a$, by, 
\bea
\LH &=& \frac{g^2}{2a}\sum_{\boldx,i}
\LE^\alpha_i(\boldx)^2 + \frac{2N}{a g^2} \sum_{\boldx, i<j} P_{ij}(\boldx),
\label{kogut-susskind}
\eea
where the plaquette operator is given by,
\be
\begin{array}{c}\includegraphics{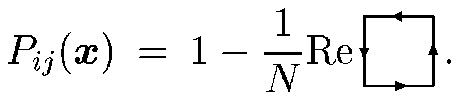}\end{array}
\label{plaquetteoperator}
\ee
%\bea
%P_{ij}(\bm{x}) &=& 1- \frac{1}{N} \Real\plaquette . 
%\label{plaquetteoperator}
%\eea

The variational parameter, $c$, is fixed by
minimising the vacuum energy density,
\be
\begin{array}{c}\includegraphics{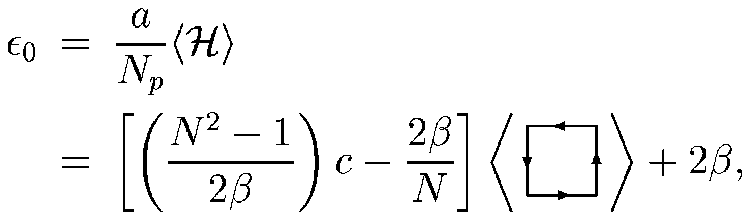}\end{array}
\label{epsilon}
\ee
%\bea
%\epsilon_0 &=& \frac{a}{N_p} \langle \LH \rangle \nn\\
%&=& \left[\left(\frac{N^2-1}{2\beta}\right) c - \frac{2\beta}
%{N}
%\right]\left\langle \plaquette \right\rangle + 2\beta,
%\label{epsilon}
%\eea
as was done for the case of 2+1 dimensions in \rcite{thesis-paper1}. 
Here $\beta = N/g^2$ and $N_p$ is the number of plaquettes on the
lattice. In 3+1
dimensions, however, the expression for the plaquette expectation value in terms of the
variational parameter is significantly more complicated than for the
case of 2+1 dimensions. Having fixed
the variational parameter we then construct a small basis of states,
with each state fitting on a single cube, and minimise the glueball mass over this
basis. The process of minimising the massgap follows precisely
\rcite{thesis-paper1}. The key difficulty of working in 3+1 dimensions is the calculation
of the required integrals. In the next section we explain how an analytic approximation to these integrals can be obtained. 

\subsection{SU(\bf{\emph{N}}) integrals in 3+1 dimensions}

The analytic techniques used in \rcite{thesis-paper1} rely on
the fact that the transformation from link to
plaquette variables has unit Jacobian in 2+1 dimensions. Batrouni has
calculated the Jacobian for arbitrary numbers of
dimensions~\cite{Batrouni:1982bg,Batrouni:1983ch}. 
For a lattice consisting of a single cube, the result of
\rcite{Batrouni:1982bg} is
\bea
J &=& \delta(P_1 P_2 P_3 P_4 P_5 P_6 -1) \nn\\
&=&  \sum_{r}\frac{1}{d_r^4}
\prod_{i=1}^{6}\chi_{r}(P_i), \label{secondline}
\eea   
where $P_1,\ldots ,P_6$ are the six plaquette variables on the single
cube and the sum is over all characters, $\chi_r$, of SU($N$). $d_r$
denotes the dimension of the character, $\chi_r$. The
second line is simply a character expansion of the first line. 
In the variational study of glueball masses, we need to 
calculate the integrals of overlapping trace variables on a single
cube. It is always possible to reduce these integrals to integrals
involving non-overlapping trace variables using the orthogonality
properties of the characters. For example, consider the expectation
value, on a single cube, of a
twice covered bent rectangle, with respect to the one-plaquette trial
state of \eqn{oneplaquette},
\be
\begin{array}{c}\includegraphics{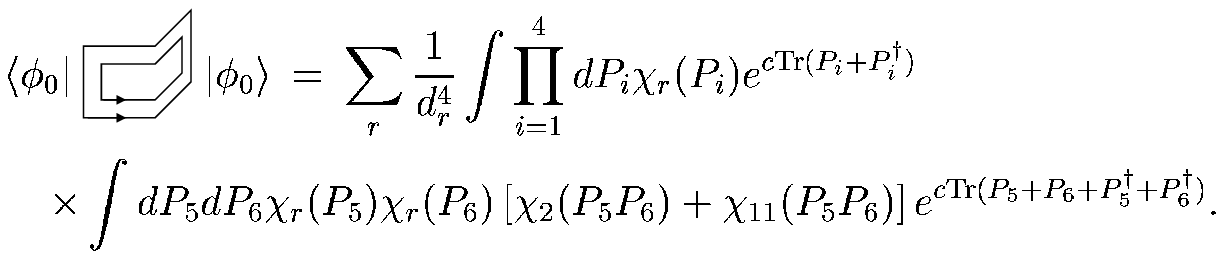}\end{array}
\label{3dexample}
\ee
%\bea
%\langle \phi_0|
%\begin{array}{c}\includegraphics[width=1.125cm]{1.eps}\end{array}|\phi_0\rangle
%&=&
%\sum_r\frac{1}{d_r^4} \int \prod_{i=1}^4 dP_i \chi_r(P_i) e^{c
%\Tr(P_i+P^\dagger_i)}\nn\\
%&&\hspace{-3cm}\times \int dP_5 dP_6  \chi_r(P_5) \chi_r(P_6) \left[\chi_2(P_5 P_6)+\chi_{11}(P_5 P_6) \right]e^{c \Tr(P_5+P_6+P^\dagger_5+P^\dagger_6)}.
%\label{3dexample}
%\eea
To proceed with this integral we need to perform character
expansions. The orthogonality of the characters can then be used to
write characters over two plaquettes in terms of one plaquette
characters. To demonstrate how this is done we consider the character expansion of one of the
integrals in \eqn{3dexample},
\bea
\int dP \chi_r(P) \chi_2(P P') e^{c \Tr(P +P^\dagger)} &=&
\sum_{r'} c_{r'} \int dP \chi_{r'}(P)  \chi_2(P P').
\eea
Here the $c_{r'}$ are given by
\bea
c_{r'} &=& \int dP \chi_{r'}(P)\chi_r(P) e^{c \Tr(P +P^\dagger)}.
\eea
Making use of the orthogonality property of characters given by, 
\bea
\int dU_p
\chi_{r'}(U_p V) \chi_{r}(U_p) = \frac{1}{d_r} \delta_{r'r} \chi_r(V),
\label{charorthog}
\eea
we obtain
\bea
\int dP \chi_r(P) \chi_2(P P') e^{c \Tr(P +P^\dagger)} &=&
\frac{c_2}{d_2} \chi_{2}(P)  \chi_2(P P').
\eea
Making use of this, and proceeding similarly for the analogous integral
involving $\chi_{11}(P_5 P_6)$, we can reduce \eqn{3dexample} to
\be
\begin{array}{c}\includegraphics{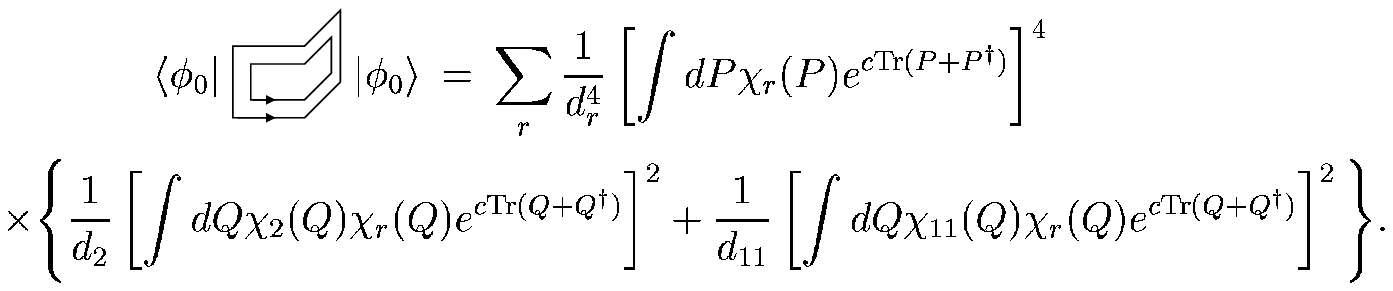}\end{array}
\ee
%\bea
%\langle \phi_0|
%\begin{array}{c}\includegraphics[width=1.125cm]{1.eps}\end{array}|\phi_0\rangle
%&=&\sum_r\frac{1}{d_r^4} \left[\int dP \chi_r(P) e^{c
%\Tr(P+P^\dagger)}\right]^4\nn\\
%&&\hspace{-5cm}\times\Bigg\{
% \frac{1}{d_2} \left[\int dQ \chi_2(Q) \chi_r(Q)
% e^{c\Tr(Q+Q^\dagger)}\right]^2 
%+ \frac{1}{d_{11}} \left[\int dQ \chi_{11}(Q) \chi_r(Q)
% e^{c\Tr(Q+Q^\dagger)}\right]^2 \Bigg\}.
%\eea
The plaquette matrix element provides a more straightforward example; 
\be
\begin{array}{c}\includegraphics{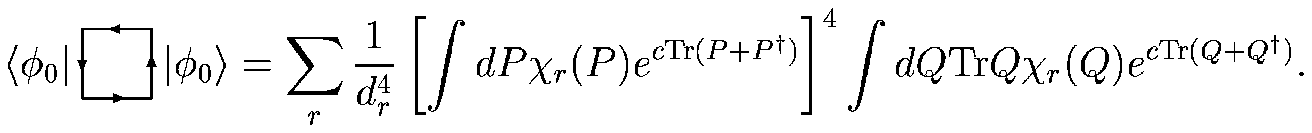}\end{array}
\label{plaqexp-3d}
\ee
%\be
%\langle \phi_0|\plaquette | \phi_0\rangle =\sum_r\frac{1}{d_r^4}\left[\int dP \chi_r(P) e^{c
%\Tr(P+P^\dagger)}\right]^4 \int dQ \Tr Q \chi_r(Q)
% e^{c\Tr(Q+Q^\dagger)}.
%\label{plaqexp-3d}
%\ee
To proceed further we need expressions for the integrals over characters that appear in each matrix element of interest. We define these character integrals 
generically by
\bea
{\cal C}_{r_1 r_2\ldots r_n}(c) &=&  \int dP \chi_{r_1}(P)
\chi_{r_2}(P)\cdots \chi_{r_n}(P) e^{c
\Tr(P+P^\dagger)}.
\eea
All SU($N$) integrals encountered in the calculation of glueball masses in 
3+1 dimensions can be expressed in terms of character integrals. 
It is possible to calculate them generally, however three points need to be 
considered. Firstly, if
we are to use the machinery of \rcite{thesis-paper1},
we need to know how
to express a given character in terms of trace variables. Secondly,
as the dimension of the gauge group 
increases the number of independent trace variables increases rapidly. 
In order to contain the number of integrals required for a
given order of approximation, it is necessary to express all high
order trace variables in terms of a basis of lower order
ones. Finally, the collection of SU($N$) integrals
presented in \rcite{thesis-paper1} must be extended. We now address each of
these points in turn.

The problem of expressing the characters of SU($N$) in terms of trace
variables was solved by Bars~\cite{Bars:1980yy} who showed that for SU($N$),
\bea
\chi_n(U) = \sum_{k_1,\ldots,k_n} \delta\left(\sum_{i=1}^n i k_i -
n\right)\prod_{j=1}^n \frac{1}{k_j! j^{k_j}} (\Tr U^j)^{k_j}. 
\label{singlebox}
\eea   
General characters can then be expressed in terms of $\chi_n(U)$ using
standard techniques from group theory~\cite{Weyl:1946},
\bea
\chi_{r_1 r_2 \cdots r_{N-1}}(U) = \det\left[ \chi_{r_j+i-j}(U)
\right]_{1\le i,j \le N-1}.
\label{chareqn}
\eea
Here the quantities inside the determinant are to be interpreted as
the $(i,j)$-th entry of an $N\times N$ matrix. 
For a given character, \eqn{chareqn} produces a multinomial of
traces of different powers of $U$. However, not all of these 
trace variables are
independent. The so called Mandelstam constraints define the
relationship between dependent trace variables. For SU(2) and SU(3) we have
\bea
\begin{array}{rcll}
\Tr U &=& \Tr U^\dagger & \forall U \in {\rm SU}(2)\quad {\rm and} \\
\Tr (U^2) &=& (\Tr U)^2 - 2 \Tr(U^\dagger) & \forall U \in {\rm SU}(3).
\end{array}
\eea  
Similar relations can be obtained to express all trace variables in
terms of $\Tr U$ for SU(2), and for SU(3) in terms of $\Tr U$ and $\Tr
U^\dagger$. To reduce the calculation of character integrals to a manageable size, we require
expressions for high power trace variables in terms
of a minimal set of lower order  trace variables. To do this we
proceed as follows.
We start with an alternative expression of $\det U = 1$, satisfied for all
SU($N$) matrices $U$,
\bea
\varepsilon_{i_1 i_2 \ldots i_{N}} U_{i_1 j_1} U_{i_2 j_2} \cdots
U_{i_N j_N} = \varepsilon_{j_1 j_2\ldots j_N}.
\label{unimod}
\eea
Here the colour indices of the group elements have been made
explicit and all repeated indices are summed over. $\varepsilon_{i_1\ldots i_n}$ is the totally antisymmetric
Levi-Civita symbol defined to be 1 if $\{i_1,\ldots,i_n\}$ is an even
permutation of $\{1,2,\ldots,n\}$, $-1$ if it is an odd permutation
and 0 otherwise (i.e. if an index is repeated). 
Multiplying Levi-Civita symbols produces sums of products
of delta functions, the precise form of which depends on how many
pairs of indices are contracted. A standard result from differential
geometry will be useful here,
\bea
\varepsilon_{a_1\ldots a_n}\varepsilon_{b_1\ldots b_n} = \det
\left(\delta_{a_i b_j}\right)_{1\le i,j\le n}.
\label{gendelta}
\eea
Multiplying both sides of \eqn{unimod} by $U^\dagger_{j_N j_{N+1}}$ and
contracting over repeated indices gives
\bea
\varepsilon_{i_1 i_2 \ldots i_{N-1}j_{N+1}} U_{i_1 j_1} U_{i_2 j_2}
\cdots U_{i_{N-1} j_{N-1}}
 &=& \varepsilon_{j_1 j_2\ldots j_N} U^\dagger_{j_N j_{N+1}} \nn\\
\varepsilon_{l_1 l_2 \ldots l_{N-1} i_N}\varepsilon_{i_1 i_2 \ldots i_{N}}
U_{i_1 j_1} U_{i_2 j_2}\cdots U_{i_{N-1} j_{N-1}} &=& \varepsilon_{l_1 l_2 \ldots l_{N-1} i_N} \varepsilon_{j_1 j_2\ldots j_N} U^\dagger_{j_N i_{N}}.
\label{unimod2}
\eea
In the last line we have renamed dummy indices and multiplied both
sides by a Levi-Civita symbol. Making use of \eqn{gendelta} in
\eqn{unimod2} produces a product of delta functions on each side of
the equation. By introducing appropriate combinations of delta
functions and contracting over repeated indices we can construct trace
identities as we please. We find that identities useful in our context
are produced by the introduction of delta functions and an additional
matrix, $A$, which is not necessarily an element of SU($N$), 
into \eqn{unimod2} as follows:
\be
\varepsilon_{k_1\cdots
k_{N-1} i_N} \varepsilon_{i_1\cdots i_N} U_{i_1 j_1}A_{j_1 k_1}  \prod_{m=2}^{N-1} \delta_{k_m
j_m} U_{i_m j_m} =  \varepsilon_{k_1\cdots
k_{N-1} i_N} \varepsilon_{j_1\cdots j_N}  A_{j_1 k_1}U^\dagger_{j_N i_N} \prod_{m=2}^{N-1} \delta_{k_m
j_m}.
\label{identgen}
\ee  
Here we have again renamed dummy indices. Some examples of 
SU($N$) identities obtained from \eqn{identgen} valid for all $N \times N$
matrices $A$ are as follows:
\bea
-\Tr U  \Tr(U A) + \Tr(U^2 A) = -\Tr A \Tr U^\dagger + \Tr(U^\dagger
 A) \quad &\forall\!\!\!& U\in {\rm SU}(3)  \nn\\
 -(\Tr U)^2 \Tr(U A) + \Tr(U A)\Tr(U^2) + 2 \Tr U \Tr(U^2 A) - 2
 \Tr(U^3 A)\nn\\
 =  -2 \Tr A \Tr U^\dagger + 2 \Tr(U^\dagger A) \quad
&\forall\!\!\! & U\in {\rm SU}(4).
\eea
We require formulae for $\Tr (U^n)$ in terms of lower order trace
variables for SU($N$).  Such formulae, known commonly as Mandelstam
constraints, are obtained by setting $A =
U^{n-N+1}$ in \eqn{identgen}. In Appendix~\ref{mandelstamconstraints} 
we list the Mandelstam
constraints obtained in this way up to SU(8). We see that for SU($N$)
it is possible to express all characters in terms of $N-1$ trace
variables. In what follows we will choose to express the general
SU($N$) characters in terms of the set of trace variables $\{\Tr U^\dagger, \Tr U, \Tr U^2,\ldots,
\Tr U^{N-2}\}$. It may well be possible to reduce the size of this
set. This would indeed improve the efficiency of our
technique. However, no
effort has been made to do this at this stage.

The procedure for calculating the general character integral ${\cal
C}_{r_1\ldots r_n}(c)$ is then as follows. We first express the
characters $\chi_{r_1},\ldots,\chi_{r_n}$ in terms of trace variables
using \eqns{singlebox}{chareqn}. We then simplify these
expressions using the Mandelstam constraints of Appendix~\ref{mandelstamconstraints}. The
final step is to perform the integrals over trace variables, which is
an increasingly non-trivial task as one increases the dimension of the
gauge group. For instance, the general character integral
 for SU($N$) involves integrals over powers of the trace variables
$\Tr U^\dagger ,\Tr U, \Tr U^2,\ldots \Tr U^{N-2} $. To proceed we need the following integral,
\bea
{\cal T}_{q_1\ldots q_k}^{p_1\ldots p_k}(c)= \int_{{\rm SU}(N)} dU (\Tr U^{q_1}
)^{p_1}(\Tr U^{q_2} )^{p_2}\cdots  (\Tr U^{q_k} )^{p_{k}}
e^{c\Tr (U+U^\dagger)}.
\eea
The cases of interest to us here for SU($N$) are described by,
$T_{-1,1,2,\ldots, N-2}^{p_1,\ldots,p_{N-1}}(c)$. 
To calculate this integral we need to extend the work of
\rcite{thesis-paper1} and consider the generating function,
\bea
G_{q_1 \ldots q_k}(c,\gamma_1,\ldots,\gamma_k) = 
\int_{{\rm SU}(N)} dU e^{c\Tr(U+U^\dagger)+ \sum_{i=1}^{k} \gamma_i 
\Tr (U^{q_i})}. 
\eea
Following the procedure of \rcite{thesis-paper1} we obtain,
\bea
G_{q_1 \ldots q_k}(c,\gamma_1,\ldots,\gamma_k) &=&
\sum_{l=-\infty}^\infty \det\left[ 
\lambda_{l+j-i,q_1,\ldots,q_k}(c,\gamma_1,\ldots,\gamma_k)
\right]_{1\le i,j \le N},
\label{gp1p2}
\eea
with,
\bea
\lambda_{m,q_1,\ldots,q_k}(c, \gamma_1, \ldots,\gamma_k) &=&
\sum_{s_1,s_2,\ldots,s_k = 0}^{\infty} \frac{\gamma_1^{s_1} \cdots
\gamma_k^{s_k}}{s_1! \cdots s_k!} I_{m+s_1 q_1 +\cdots +s_k q_k}( 2 c).
\eea
We then obtain an expression for ${\cal T}_{q_1\ldots q_k}^{p_1\ldots
p_{k}}(c)$ by differentiating $G_{q_1\ldots
q_{k}}$ appropriately;
\bea
{\cal T}^{p_1\ldots p_{k}}_{q_1\ldots q_{k}}
(c) &=& \frac{\partial^{p_1+\cdots+p_{k}}}{\partial
\gamma_{1}^{p_{1}} \cdots  \partial\gamma_{k}^{p_{k}}}
G_{q_1 \ldots q_{k}}(c,\gamma_1,\ldots,\gamma_k)
\Bigg|_{\gamma_{1} =\cdots =\gamma_{k} =0}.
\label{genfunc-3d}
\eea

\subsection{The variational ground state}

In this section we fix the variational ground state following the
usual procedure of minimising the unimproved vacuum energy density
given by \eqn{epsilon}. This equation is independent
of the number of dimensions. The dimensionality of the lattice arises
at the stage of calculating the plaquette expectation value. 

Making use of \eqn{plaqexp-3d}, the variational parameter can be fixed as
a function of $\beta$ for the one-cube lattice. In practice, the
character sum in \eqn{plaqexp-3d} and the infinite $l$-sum in \eqn{gp1p2} 
need to be truncated. We truncate the infinite $l$-sum at $\pm l_{max}$ and instead of summing over all SU($N$) characters, we sum over 
only those characters, $r = (r_1,r_2,\ldots,r_{N-1})$, with
$r_{max}\ge r_1\ge r_2 \ge \cdots \ge r_{N-1}$. With this truncation
scheme, memory constraints restrict calculations of the variational SU($N$) ground state to $N\le 7$, when
working with $r_{max}=2$ and $l_{max}=2$ on a desktop computer.

\begin{figure}
\centering
                       
\includegraphics[width=10cm]{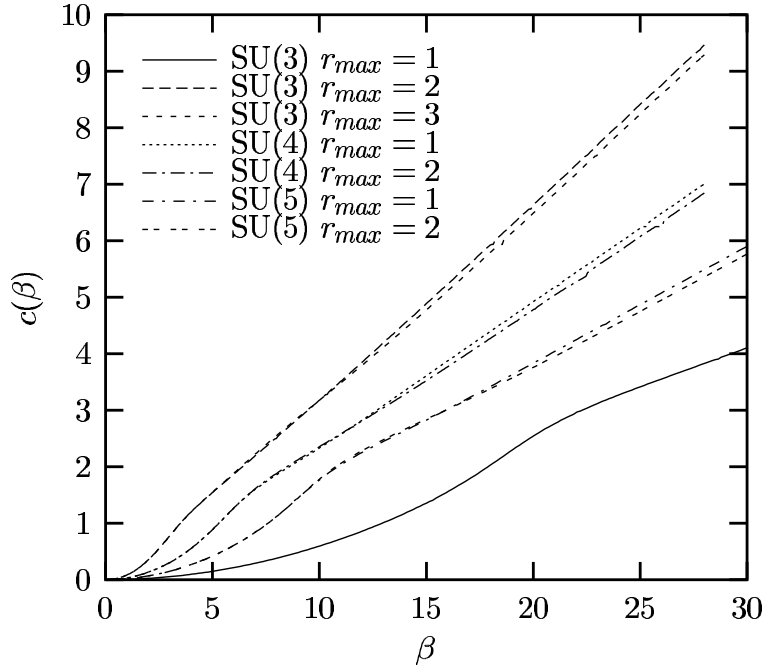}

\caption{The one cube SU($N$) variational parameter 
as a function of $\beta$ for various $N$ showing the dependence
on the character sum truncation.}
\label{c-convwithchars}  
\end{figure}

The dependence of the variational parameter on $r_{max}$ with
$l_{max}=2$ is shown
for various gauge groups in \fig{c-convwithchars}. As $r_{max}$ increases the 
variational parameter appears to
converge for each $N$ considered. Moreover the convergence appears to improve as the dimension
of the gauge group is increased. With the exception of SU(3), the $r_{max}=1$ and 2
results are indistinguishable on the range $0\le \xi\le 0.7$, where
$\xi = 1/(N g^2)$. As this
is the range of interest to us later in the paper we restrict
further calculations to $r_{max} = 1$. 

\begin{figure}
\centering
                       
\includegraphics[width=10cm]{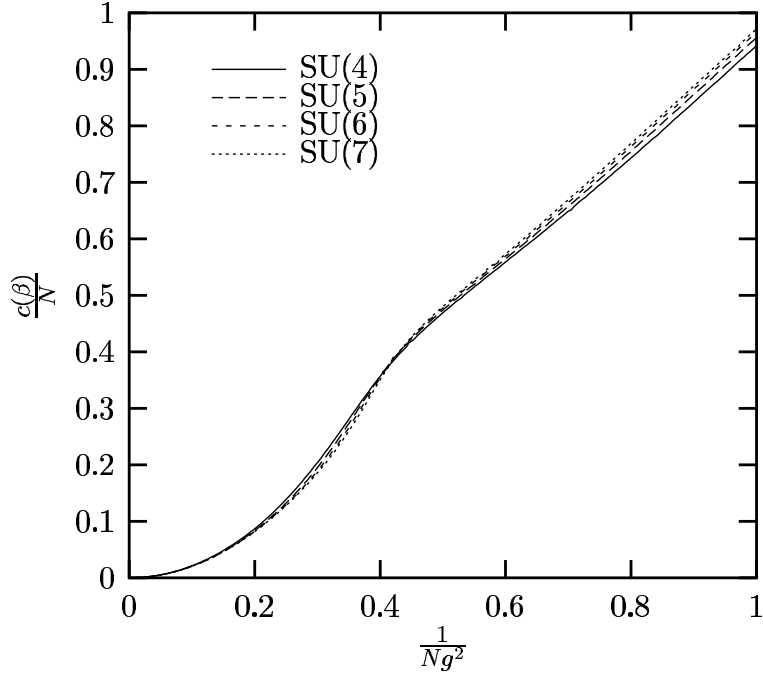}

\caption{The one cube SU($N$) variational parameter in units
of $N$ as a function of $1/(N g^2)$ for various $N$.}
\label{3+1sunvarpar}  
\end{figure}

\begin{figure}
\centering
                       
\includegraphics[width=10cm]{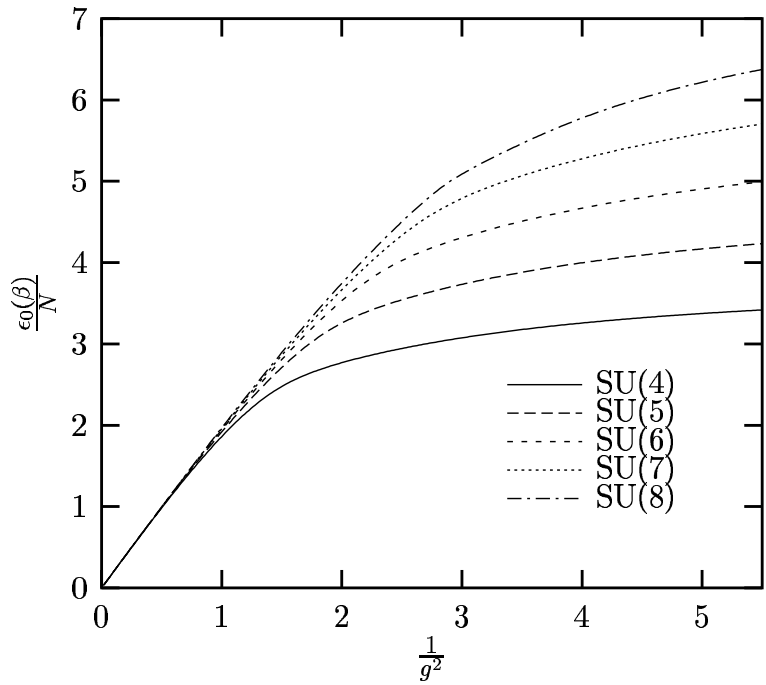}

\caption{The one cube SU($N$) energy density in units
of $N$ as a function of $1/g^2$ for various $N$.}
\label{3+1sunedens}  
\end{figure}

The SU($N$) variational parameters on the one-cube lattice with
$r_{max}=1$ and $l_{max}=2$ are shown for various $N$ in
\fig{3+1sunvarpar}. The results do not differ greatly. The corresponding variational energy
densities are shown in \fig{3+1sunedens}.

\subsection{Expressions for the glueball mass}

We follow precisely the method described in \rcite{thesis-paper1} for
the calculation of 3+1 dimensional glueball masses on a single cube. 
Here, however, we choose a different basis of
states to minimise over. Instead of rectangular loops, we use states
which fit in a single cube. We start with a basis of two states, the
plaquettes and the bent rectangles,
\bea
B &=& \{ |1\rangle ,|2\rangle \},
\label{basis}
\eea
with 
\bea
|i\rangle &=&  \sum_{\bms{x}} \left[F_i(x)-\langle F_i(x)
\rangle\right] |\phi_0 \rangle
\eea 
and 
\be
\begin{array}{c}\includegraphics{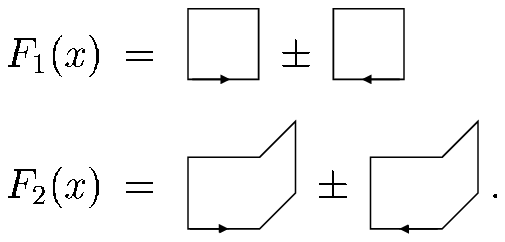}\end{array}
\ee
%\bea
%F_1(x) &=& \begin{array}{c}\includegraphics[width=0.75cm]{30.eps}\end{array}
%\pm \begin{array}{c}\includegraphics[width=0.75cm]{31.eps}\end{array}\nn\\
%F_2(x)
%&=&\begin{array}{c}\includegraphics[width=1.125cm]{9.eps}\end{array}
%\pm \begin{array}{c}\includegraphics[width=1.125cm]{32.eps}\end{array}.
%\eea 
In 3+1 dimensions, the ``+'' sign corresponds to the $0^{++}$ state and 
the ``$-$'' sign corresponds to the $1^{+-}$ state~\cite{Hamer:1989qm}. 
In order to calculate the glueball masses, following
Arisue~\cite{Arisue:1983tt}, we need expressions for the matrix
elements $N^C_{ii'}$ and $D^C_{ii'}$~\cite{thesis-paper1} defined by,
\bea
N^C_{l l'} &=& \frac{1}{N_p}\langle l|\tilde{H} - E_0|l'\rangle
\label{Nl'l}
\eea
and
\bea
D^C_{l l'} &=& \frac{1}{N_p}\langle l|l'\rangle 
= \sum_{\boldx}\left[\langle F^\dagger_l(\boldx) F_{l'}(\boldsymbol{0})
\rangle  - \langle F_l(\boldx) \rangle^\ast 
\langle F_{l'}(\boldsymbol{0}) \rangle\right].     
\label{d}
\eea
 Here the superscript, $C$, denotes the charge conjugation eigenvalue, $C=\pm 1$, of the state in question. 
Following Arisue~\cite{Arisue:1983tt} 
we have the following reduction (which was
 generalised to improved Hamiltonians in \rcite{thesis-paper1}):
\bea
N^C_{l l'} &=& -\frac{g^2}{2a}\sum_{i,\boldx}\sum_{\boldx'}\left\langle \left[E^\alpha_i(\boldx),F^\dagger_{l}(\boldx') \right]\left[E^\alpha_i(\boldx),F_{l'}(\boldsymbol{0})\right]\right\rangle.
\label{N}
\eea 
After carefully counting the number of possible overlaps between different loops we arrive at:
\be
\begin{array}{c}\includegraphics{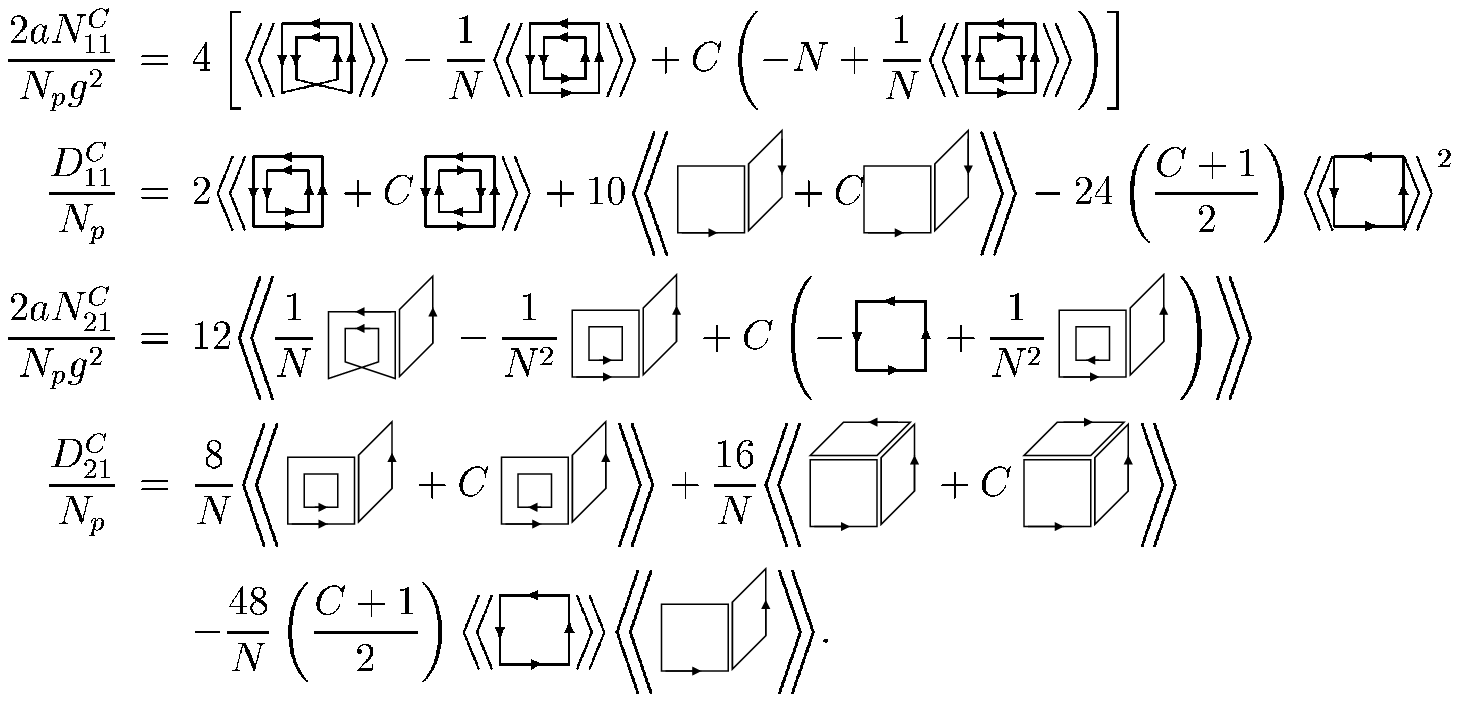}\end{array}
\label{mes-3+1}\ee
%\bea
%\frac{2aN^C_{11}}{N_p g^2} &=&
%4 \left[ \flangle\grapha\frangle
%-\frac{1}{N}\flangle \graphd \frangle+C\left(- N +\frac{1}{N}\flangle \graphdd\frangle\right)
%\right]  \nn\\
% \frac{D^C_{11}}{N_p} &=&
%2 \flangle\graphd + C \graphdd \frangle
%+10 \bflangle \begin{array}{c}\includegraphics[width=1.125cm]{21.eps}\end{array} 
%\!\!+ C \!\!\begin{array}{c}\includegraphics[width=1.125cm]{21.eps}\end{array}\bfrangle
%-24\left(\frac{C+1}{2}\right) \flangle\!\plaquette\! \frangle^2 \nn\\
%\frac{2aN^C_{21}}{N_p g^2} &=& 12 \bflangle
%\frac{1}{N} \begin{array}{c}\includegraphics[width=1.125cm]{29.eps}\end{array}
%-\frac{1}{N^2} \begin{array}{c}\includegraphics[width=1.125cm]{16.eps}\end{array}
%+C\left(-
%\plaquette
% +\frac{1}{N^2}
% \begin{array}{c}\includegraphics[width=1.125cm]{17.eps}\end{array}\right)\bfrangle\nn\\
% \frac{D^C_{21}}{N_p} &=&
% \frac{8}{N}\bflangle\begin{array}{c}\includegraphics[width=1.125cm]{16.eps}\end{array}+C\begin{array}{c}\includegraphics[width=1.125cm]{17.eps}\end{array}\bfrangle+\frac{16}{N}\bflangle\begin{array}{c}\includegraphics[width=1.125cm]{14.eps}\end{array}+
% C
% \begin{array}{c}\includegraphics[width=1.125cm]{15.eps}\end{array}\bfrangle
% \nn\\
%&& - \frac{48}{N}\left(\frac{C+1}{2}\right) \flangle \plaquette
% \frangle \bflangle
% \begin{array}{c}\includegraphics[width=1.125cm]{20.eps}\end{array}\bfrangle .\label{mes-3+1}
%\eea
Here we have introduced the notation
\bea
\langle \!\langle O \rangle\!\rangle = \langle \phi_0 | O |\phi_0 \rangle. 
\eea
The combinatorics which lead to the coefficients in the matrix
elements in \eqn{mes-3+1} are a result of counting the possible 
overlaps within a
single cube. The remaining matrix elements $N_{22}$ and $D_{22}$ can
be calculated similarly, resulting in
\be
\begin{array}{c}\includegraphics{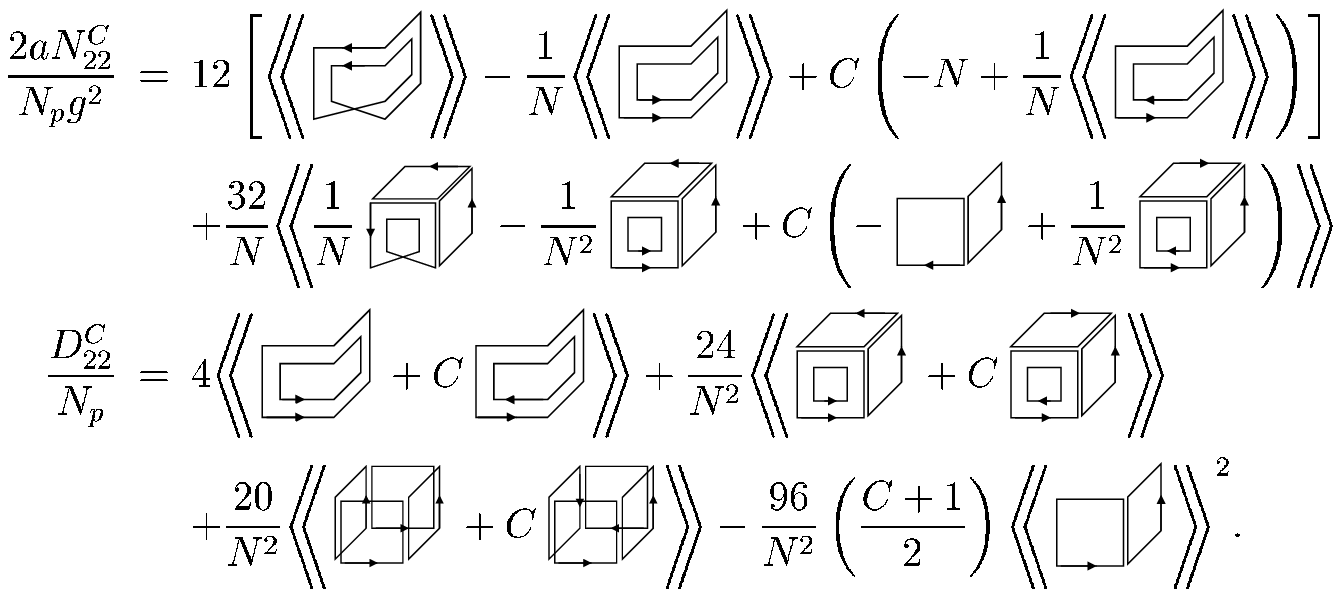}\end{array}
\ee
%\bea
%\frac{2aN^C_{22}}{N_p g^2} &=& 12\left[\bflangle
% \begin{array}{c}\includegraphics[width=1.125cm]{26.eps}\end{array}\bfrangle
%-\frac{1}{N}\bflangle\begin{array}{c}\includegraphics[width=1.125cm]{1.eps}\end{array}\bfrangle 
%+C\left(-N+\frac{1}{N}
% \bflangle \begin{array}{c}\includegraphics[width=1.125cm]{2.eps}\end{array}\bfrangle\right)
%\right] \nn\\
%&& + \frac{32}{N} \bflangle
%\frac{1}{N}\begin{array}{c}\includegraphics[width=1.125cm]{45.eps}\end{array}
%-
%\frac{1}{N^2}
%\begin{array}{c}\includegraphics[width=1.125cm]{18.eps}\end{array}
%+C\left(-
%  \begin{array}{c}\includegraphics[width=1.125cm]{35.eps}\end{array}
%+\frac{1}{N^2}\begin{array}{c}\includegraphics[width=1.125cm]{19.eps}\end{array}
%\right)\bfrangle \nn\\
%\frac{D^C_{22}}{N_p} &=&
%  4\bflangle \begin{array}{c}\includegraphics[width=1.125cm]{1.eps}\end{array}+C
%  \begin{array}{c}\includegraphics[width=1.125cm]{2.eps}\end{array}\bfrangle
%  + \frac{24}{N^2}\bflangle
%\begin{array}{c}\includegraphics[width=1.125cm]{18.eps}\end{array}
%  + C \begin{array}{c}\includegraphics[width=1.125cm]{19.eps}\end{array}
%\bfrangle \nn\\
%&& + \frac{20}{N^2} \bflangle \begin{array}{c}\includegraphics[width=1.125cm]{24.eps}\end{array}
%  + C \begin{array}{c}\includegraphics[width=1.125cm]{25.eps}\end{array}
%\bfrangle -\frac{96}{N^2}\left(\frac{C+1}{2}\right)\bflangle \begin{array}{c}\includegraphics[width=1.125cm]{20.eps}\end{array} \bfrangle^2.
%\eea
Having calculated $N^C_{ii'}$ and $D^C_{ii'}$, we follow the
minimisation procedure described in \rcite{thesis-paper1} to arrive at
the glueball mass, $\Delta M^{PC}$.

\subsection{Results}
\label{results-3+1}
In this section we present calculations of the $0^{++}$ (symmetric) and $1^{+-}$
(antisymmetric) glueball masses on the
one-cube lattice for SU($N$) with $4\le N\le7$. 
We first define the rescaled glueball masse, $\mu^{PC}$, corresponding
to $\Delta M^{PC}$ as follows:
\bea
 \mu^{PC} = \log( a \Delta M^{PC} \xi^{-51/121}) -\frac{51}{121}\log\left(\frac{48 \pi^2}{11}\right) +\frac{24
 \pi^2}{11} \xi .
\eea
Here, as usual, $P$ and $C$ denote the parity and charge conjugation 
eigenvalues
of the state in question. As a result of standard renormalisation group
arguments~\cite{Caswell:1974gg,Jones:1974mm}, asymptotic scaling of a
glueball mass is observed if the corresponding rescaled glueball mass
becomes constant for some range of couplings.

\begin{figure}
\centering
               
\subfigure[Calculated with only plaquettes in the minimisation basis] % caption for subfigure a
                     {
                         \label{S-1-0}
                         \includegraphics[width=7cm]{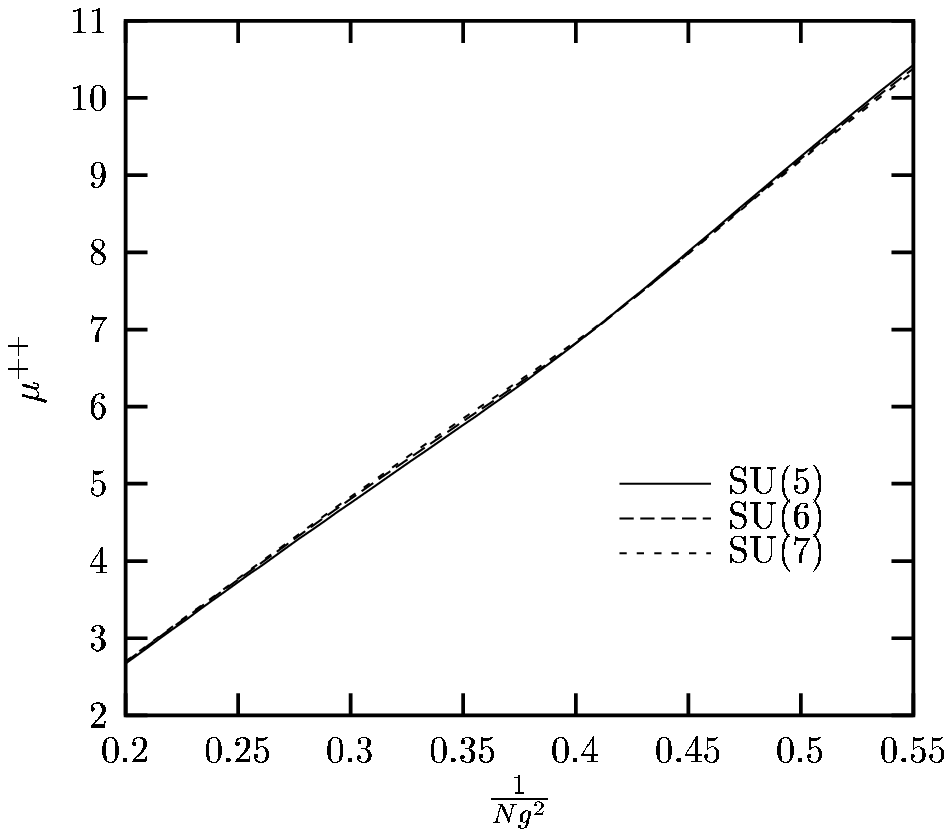}
                     }\hspace{0.25cm}
\subfigure[Calculated with both plaquettes and bent rectangles in the
                         minimisation basis] % caption for subfigure a
                     {
                         \label{S-1-1}
                         \includegraphics[width=7cm]{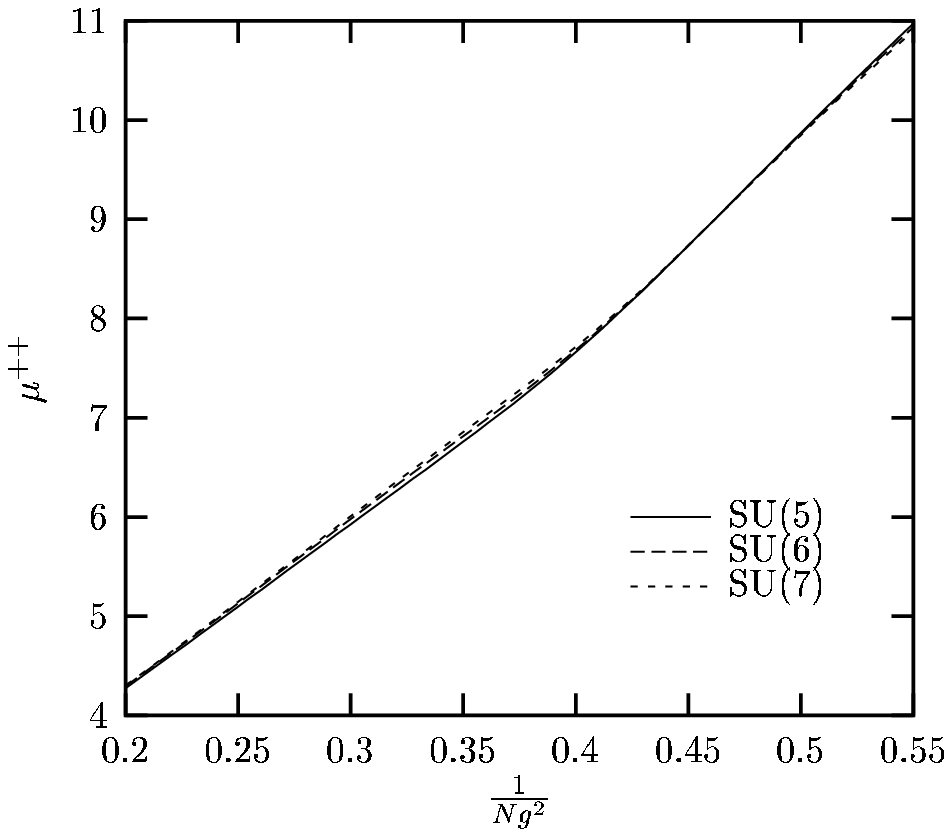}
                     }
\caption{The one cube SU($N$) $0^{++}$ rescaled glueball mass as a
function of $1/(N g^2)$ for various $N$ obtained with different
minimisation bases.}
\label{3+1sunmassgap-S}  
\end{figure}

The results
for the rescaled symmetric glueball mass are shown in
\fig{3+1sunmassgap-S}. \fig{S-1-0} shows the rescaled glueball mass
 calculated with only plaquettes in the minimisation
basis. \fig{S-1-1} shows the same quantity calculated with the
minimisation basis of \eqn{basis}.  The
aim of this exploratory study is to observe whether or not a move
towards scaling is apparent 
as the number of states in the minimisation basis is increased. This
is clearly not the case for the symmetric glueball mass on the range of
couplings explored here. There is no sign of $\mu^{++}$ becoming constant on any range of couplings within $0\le \xi \le 0.55$.

\begin{figure}
\centering
               
\subfigure[Calculated with only plaquettes in the minimisation basis] % caption for subfigure a
                     {
                         \label{AS-1-0}
                         \includegraphics[width=7cm]{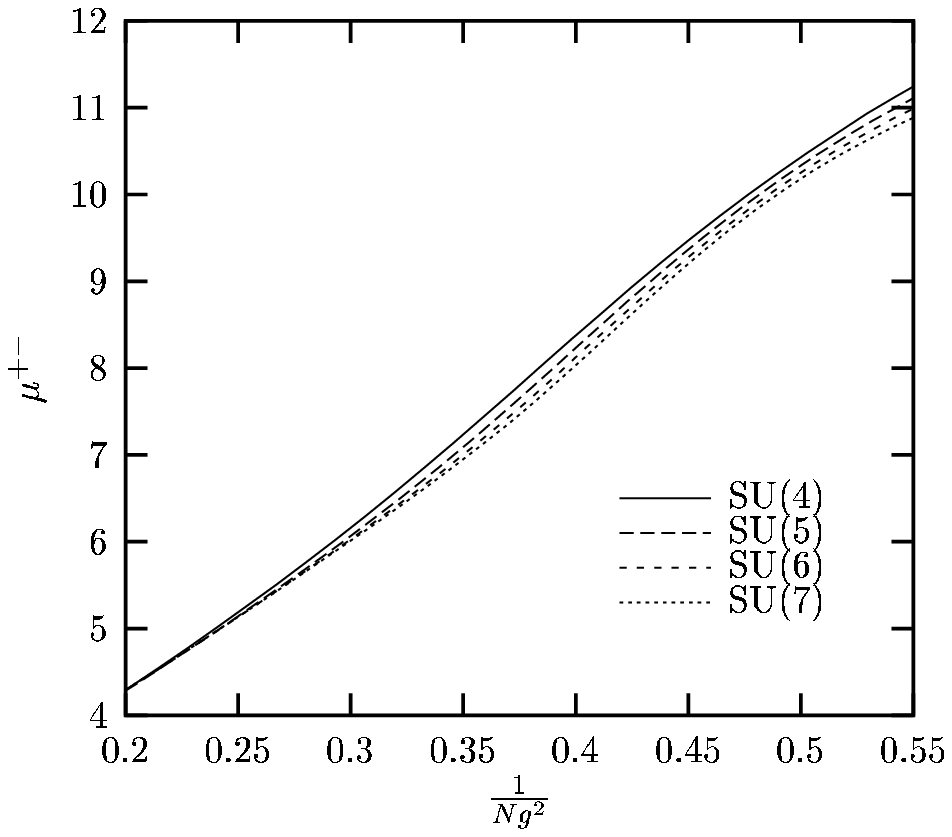}
                     }\hspace{0.25cm}
\subfigure[Calculated with both plaquettes and bent rectangles in the
                         minimisation basis] % caption for subfigure a
                     {
                         \label{AS-1-1}
                         \includegraphics[width=7cm]{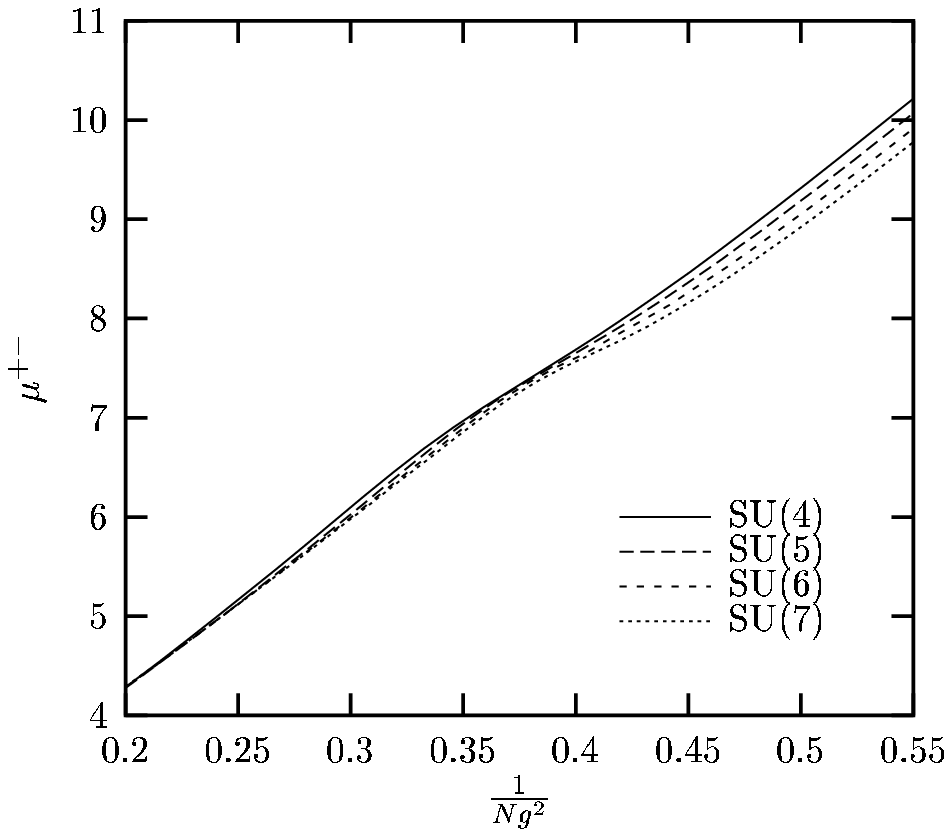}
                     }
\caption{The one cube SU($N$) $1^{+-}$ rescaled glueball mass as a
function of $1/(N g^2)$ for various $N$ obtained with different
minimisation bases.}
\label{3+1sunmassgap-AS}  
\end{figure}

The prospects are marginally better for the antisymmetric case.
The results for the rescaled antisymmetric glueball mass are displayed in
\fig{3+1sunmassgap-AS}. A move towards scaling is observed,
most clearly for the $N=7$ case, near $\xi=0.38$. This matches the region
of couplings for which Chin, Long and Robson observed scaling of the $0^{++}$ glueball mass 
for $N=5$ and 6 on a $6^3$ lattice 
using only plaquettes in their
minimisation basis~\cite{Chin:1986fe}.

\subsection{String tension}

If the exploratory study presented here was to be extended to larger
lattices, at some stage it would be useful to compute the string
tension, $\sigma$. It is common for calculations in Lagrangian LGT to
express results for masses in units of $\sqrt{\sigma}$. This allows
masses calculated on the lattice to be expressed in $MeV$ since the string tension can be 
calculated in $MeV$
from the decay of heavy quarkonia for example.   In the
Hamiltonian formulation, precise calculations of the string tension have only
been performed in the strong coupling regime. Variational
estimates, at least for SU(2)~\cite{Arisue:1983tt}, have not exhibited asymptotic scaling
when making use of the one plaquette ground state of \eqn{oneplaquette}. It is possible that the situation may be improved for higher dimensional gauge groups  but this has not yet been tested.

In this section we calculate the symmetric and antisymmetric
glueball masses in units of $\sqrt{\sigma}$. For the string tension, since
reliable variational results are not available, we use the strong
coupling expansions of Kogut and Shigemitsu~\cite{Kogut:1980pm}. These
are available for SU(5) and SU(6), as well as SU(2) and SU(3)
which we do not consider here.

Since the square root of the string tension has units of mass, the ratio of a 
mass to the string tension is constant in a
scaling region. Again, with the crude model presented here, we do not expect
to observe scaling. We seek only an indication of an approach to
scaling. Such behaviour would warrant further study. 

\begin{figure}
\centering
               
\subfigure[Lowest energy state] % caption for subfigure a
                     {
                         \label{S-1-both}
                         \includegraphics[width=7cm]{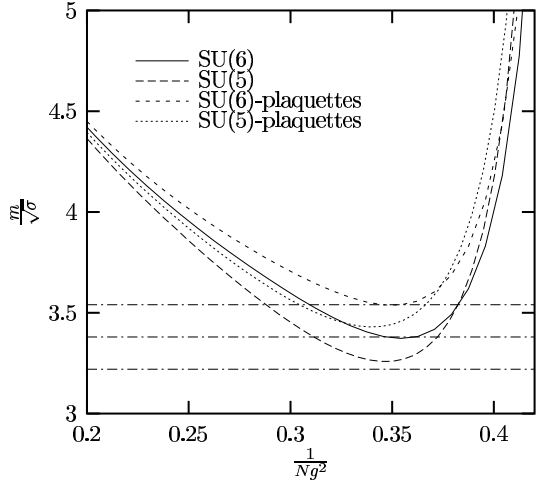}
                     }\hspace{0.25cm}
\subfigure[Second lowest energy state] % caption for subfigure a
                     {
                         \label{S-2}
                         \includegraphics[width=7cm]{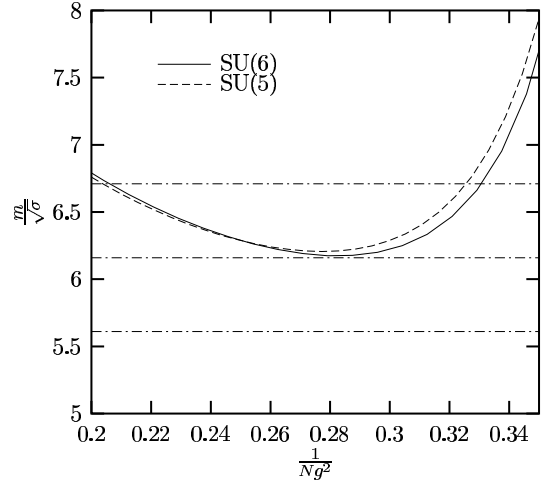}
                     }
\caption{The one cube SU(5) and SU(6) $0^{++}$ 
glueball masses in units of $\sqrt{\sigma}$ as functions of
$1/(N g^2)$. The horizontal lines indicate the result and error bars
of the SU(5) $0^{++}$ calculation of Lucini and Teper~\cite{Lucini:2001ej}.}
\label{3+1-S-ST}  
\end{figure}

 The results for the symmetric states are
shown in \fig{3+1-S-ST}. Since we perform calculations with two
states in the minimisation basis, the two lowest mass states are
accessible. In \fig{S-1-both} the results for the lowest mass
state are shown. Calculations with different minimisation bases are
shown, with the ``SU$(N)$-plaquettes'' label indicating that only  plaquette states are used. 
 For each case the glueball mass has a
local minimum in the range $0.33\le\xi\le 0.37$.  
The various results do not differ greatly. There is no sign of
improved scaling behaviour when either bent rectangles are included or
$N$ is increased. The effect of including bent rectangles is to lower
the local minimum.

The second lowest glueball mass is shown in
\fig{S-2}. Again, no improvement in scaling behaviour is seen
as $N$ is increased. In \fig{3+1-S-ST}
the horizontal lines indicate the 3+1 dimensional SU(5) calculations (and
error bars) of Lucini and Teper~\cite{Lucini:2001ej}. Interestingly,
both masses calculated here with a simplistic model, have minima that lie
within the error bars.

\begin{figure}
\centering
               
\subfigure[Lowest energy state] % caption for subfigure a
                     {
                         \label{AS-1-both}
                         \includegraphics[width=7cm]{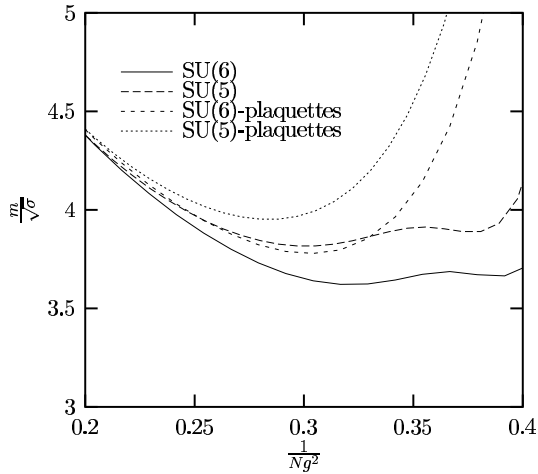}
                     }\hspace{0.25cm}
\subfigure[Second lowest energy state] % caption for subfigure a
                     {
                         \label{AS-2}
                         \includegraphics[width=7cm]{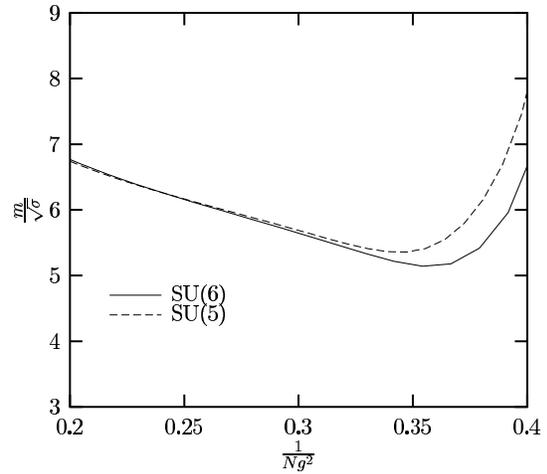}
                     }
\caption{The one cube SU(5) and SU(6) $1^{+-}$ glueball masses in units of $\sqrt{\sigma}$ as functions of
$1/(N g^2)$.}
\label{3+1-AS-ST}  
\end{figure}

We now move on to the antisymmetric states. 
The results for the antisymmtric glueball masses in units of
$\sqrt{\sigma}$ are shown in \fig{3+1-AS-ST}. The results are
more promising than the symmetric case. We see an improved approach to
scaling when bent rectangles are included in the minimisation basis
for the lowest glueball mass. For this case the glueball mass shows
promising signs of becoming constant in the ranges  $0.32\le \xi \le 0.39$ for
SU(6) and $0.3\le \xi \le 0.37$ for SU(5). Only
marginal improvement in scaling behaviour is evident as $N$ is
increased from 5 to 6. For each of these cases
no data is available for comparison to our knowledge. Such promising
signs are not apparent in the second lowest glueball mass in this
sector as seen in \fig{AS-2}.

It would be interesting to perform analogous calculations for SU(7),
for which the most promising results were displayed in
\sect{results-3+1}. However strong coupling expansions for the
 SU(7) string tension are not available in \rcite{Kogut:1980pm} and have not been published elsewhere to
our knowledge.

%Discuss the problem with working in 3+1 dimensions. Why is it so
%vastly different to 2+1 dimensions? Discuss what has been done, in
%particular, how the problem has been avoided. Discuss why the
%plaquette expansion technique will not work. Discuss what can be done
%within a plaquette approach.

\section{Future Work}
\label{futurework}
In this paper we have studied the variational $0^{++}$ and $1^{+-}$ glueball masses
 on a single cube in 3+1 dimensions. 
Our intention was to determine the viability of using analytic
 Hamiltonian methods in 3+1 dimensional glueball mass
 calculations. With such a crude small volume approximation and a
 minimisation basis containing only two states, the observation of
 asymptotic scaling was not expected. Promising signs of an approach
 to asymptotic scaling was displayed by the $1^{+-}$ glueball mass, as $N$ was increased, at couplings close to the 
scaling window observed by a comparable, but larger volume, study by Chin, Long and Robson~\cite{Chin:1986fe}. Interesting results were also observed for calculations of the glueball masses in units of the string tension. With no variational results for the string tension available, the strong coupling results of Kogut and Shigemitsu~\cite{Kogut:1980pm} were used. Interestingly, the mass of both SU(5) $0^{++}$ glueball states calculated had minima which were consistent with the Lagrangian calculation of Lucini and Teper~\cite{Lucini:2001ej}. Better scaling behaviour was exhibited by the SU(5) and SU(6) $1^{+-}$ states, although no alternative calculations are 
available for comparison to our knowledge. For the lowest mass $1^{+-}$ states calculated, the scaling behaviour was improved by increasing the number of states in the minimisation basis and the dimension of the gauge group. The promising results observed in this paper warrant further study on larger lattices, with additional states in the minimisation basis.

To extend this calculation to larger lattices and minimisation bases, two challenges will be faced. Firstly, the correct implementation of the Bianchi identity, within our plaquette based analytic approach, forces the number of integration variables to grow quickly with the volume. New approaches to handling the general character integrals may need to be developed in order to avoid the inevitable memory restrictions. Secondly, great care will need to be taken in the counting of overlaps between different states. This process could, in principle, be automated using techniques from graph theory and symbolic programming.

It would also be of interest to extend the calculation presented here to larger $N$.
To do this a more efficient technique for handling the general character integrals would need to be developed to minimise the memory demands of the calculation. 
The following is just one possibility. It is likely, that by expressing the products of characters appearing in the character integrals as sums of characters, one could reduce the memory requirements of the calculation drastically. The general character integrals could then be expressed in terms of integrals of the form,
\bea
G_r(c) &=& \int_{{\rm SU}(N)} dU \chi_r(U^\dagger) 
e^{c \Tr(U + U^\dagger)},
\eea
where $r = (r_1,r_2,\ldots r_{N-1})$ labels a representation of SU($N$).
Having completed the calculations presented in this paper, it was discovered that $G_r(c)$ can be handled using the techniques of \rcite{thesis-paper1}, with the result,
\bea
G_r(c) &=& \sum_{l=-\infty}^\infty \det \left[I_{r_i + l+ j -i}(2 c) \right]_{1\le i,j\le N}.
\eea 
The final stage of this improvement would be to find a convenient way
to express the products of characters appearing in the calculation as a sum of characters. This would be possible with the symbolic manipulation of Young tableaux.

\begin{acknowledgments}
We wish to acknowledge useful and interesting discussions with J.~A.~L.~McIntosh and L.~C.~L.~Hollenberg. We also thank R.~F.~Bishop for bringing the problem of Gauss' law in 3+1 dimensions to our attention.
\end{acknowledgments}
 
\appendix

\section{Mandelstam Constraints}
\label{mandelstamconstraints}
\bea
\Tr U^n &=& \Tr U^{n-3}-\Tr U^\dagger \Tr U^{n-2} + \Tr U \Tr U^{n-1}
\qquad \forall U\in \rm{SU}(3)
\eea
\bea
\Tr U^n &=& -\Tr U^{n-4}+\Tr U^\dagger \Tr U^{n-3} -\frac{1}{2} (\Tr
U)^2 \Tr U^{n-2} + \frac{1}{2} \Tr
U^2 \Tr U^{n-2} \nn\\
&& + \Tr U \Tr U^{n-1}
\qquad \forall U\in \rm{SU}(4)
\eea
\bea
\Tr U^n &=&
    \Tr U^{n-5} - \Tr U^\dagger \Tr U^{n-4} + 
      \frac{1}{6}  (\Tr U)^3 \Tr U^{n-3} - 
      \frac{1}{2} \Tr U \Tr U^2 \Tr U^{n-3} \nn\\
&&  + 
      \frac{1}{3} \Tr U^3 \Tr U^{n-3} - \frac{1}{2} (\Tr U)^2 \Tr U^{n-2} + 
      \frac{1}{2} \Tr U^2 \Tr U^{n-2} \nn\\
&& + \Tr U \Tr U^{n-1} \qquad \forall U\in \rm{SU}(5)
\eea
\bea
\Tr U^n &=& -\Tr U^{n-6} + 
      \Tr U^\dagger\Tr U^{n-5} - 
      \frac{1}{24} (\Tr U)^4\Tr U^{n-4}+ 
      \frac{1}{4} (\Tr U)^2\ \Tr U^2 \Tr U^{n-4}\nn\\
&& - 
      \frac{1}{8} (\Tr U^2)^2 \Tr U^{n-4} - 
      \frac{1}{3} \Tr U \Tr U^3 \Tr U^{n-4}+ 
      \frac{1}{4} \Tr U^4\Tr U^{n-4}+ 
      \frac{1}{6} (\Tr U)^3\Tr U^{n-3} 
\nn\\ 
&&
     - \frac{1}{2} \Tr U \Tr U^2 \Tr U^{n-3}+ 
      \frac{1}{3} \Tr U^3\Tr U^{n-3} - 
      \frac{1}{2} \Tr U^2\Tr U^{n-2} + 
      \frac{1}{2} \Tr U^2 \Tr U^{n-2} \nn\\
&& +\Tr U \Tr U^{n-1}  \qquad \forall U\in \rm{SU}(6)
\eea
\bea
 \Tr  U^n  &=& 
    \Tr U^{n-7} - \Tr U^\dagger \Tr U^{n-6} + 
      \frac{1}{120}(\Tr U)^5\Tr U^{n-5} - 
      \frac{1}{12}(\Tr U)^3\Tr U^2 \Tr U^{n-5} 
\nn\\&&+ 
      \frac{1}{8}\Tr U (\Tr U^2)^2\Tr U^{n-5} + 
      \frac{1}{6}(\Tr U)^2\Tr U^3 \Tr U^{n-5} - 
      \frac{1}{6}\Tr U^2 \Tr U^3 \Tr U^{n-5}
\nn\\&& - 
      \frac{1}{4}\Tr U \Tr U^4 \Tr U^{n-5}+ 
      \frac{1}{5}\Tr U^5 \Tr U^{n-5} - 
      \frac{1}{24}(\Tr U)^4\Tr U^{n-4} + 
      \frac{1}{4}(\Tr U)^2\Tr  U^2 \Tr U^{n-4}
\nn\\&&  - 
      \frac{1}{8}(\Tr U^2)^2\Tr U^{n-4}  - 
      \frac{1}{3}\Tr U \Tr U^3 \Tr U^{n-4}+ 
      \frac{1}{4}\Tr U^4 \Tr U^{n-4}  + 
      \frac{1}{6}(\Tr U)^3\Tr U^{n-3}
\nn\\&&  - 
      \frac{1}{2}\Tr U \Tr U^2 \Tr U^{n-3}+ 
      \frac{1}{3}\Tr U^3 \Tr U^{n-3}- 
      \frac{1}{2}(\Tr U)^2\Tr U^{n-2}  + 
      \frac{1}{2}\Tr U^2 \Tr U^{n-2}
\nn\\&&  + \Tr U \Tr U^{n-1}\qquad \forall U\in \rm{SU}(7)
\eea
\bea  
 \Tr U^n  &=& -\Tr U^{n-8}  + 
      \Tr U^\dagger \Tr U^{n-7}  - 
      \frac{1}{720}(\Tr U)^6\Tr U^{n-6}  + 
      \frac{1}{48}(\Tr U)^4\Tr U^2 \Tr U^{n-6} 
\nn\\&& - 
      \frac{1}{16}(\Tr U)^2(\Tr U^2)^2\Tr U^{n-6} + 
      \frac{1}{48}(\Tr U^2)^3\Tr U^{n-6}  - 
      \frac{1}{18}(\Tr U)^3\Tr U^3 \Tr U^{n-6}  
\nn\\&&+ 
      \frac{1}{6}\Tr U \Tr U^2 \Tr U^3 \Tr U^{n-6}  - 
      \frac{1}{18}(\Tr U^3)^2\Tr U^{n-6} + 
      \frac{1}{8}(\Tr U)^2\Tr U^4 \Tr U^{n-6}
\nn\\&&  - 
      \frac{1}{8}\Tr U^2 \Tr U^4 \Tr U^{n-6}  - 
      \frac{1}{5}\Tr U \Tr U^5 \Tr U^{n-6}+ 
      \frac{1}{6}\Tr U^6 \Tr U^{n-6}  + 
      \frac{1}{120}(\Tr U)^5\Tr U^{n-5} 
\nn\\&& - 
      \frac{1}{12}(\Tr U)^3\Tr U^2 \Tr U^{n-5}  + 
      \frac{1}{8}\Tr U (\Tr U^2)^2\Tr U^{n-5} + 
      \frac{1}{6}(\Tr U)^2\Tr U^3 \Tr U^{n-5}
\nn\\&& - 
      \frac{1}{6}\Tr U^2 \Tr U^3 \Tr U^{n-5}  - 
      \frac{1}{4}\Tr U \Tr U^4 \Tr U^{n-5}+ 
      \frac{1}{5}\Tr U^5 \Tr U^{n-5} - 
      \frac{1}{24}(\Tr U)^4\Tr U^{n-4}
\nn\\&& + 
      \frac{1}{4}(\Tr U)^2\Tr U^2 \Tr U^{n-4}  - 
      \frac{1}{8}(\Tr U^2)^2\Tr U^{n-4}  - 
      \frac{1}{3}\Tr U \Tr U^3 \Tr U^{n-4}+ 
      \frac{1}{4}\Tr U^4 \Tr U^{n-4} 
\nn\\&&+ 
      \frac{1}{6}(\Tr U)^3\Tr U^{n-3}  - 
      \frac{1}{2}\Tr U \Tr U^2 \Tr U^{n-3}  + 
      \frac{1}{3}\Tr U^3 \Tr U^{n-3}  - 
      \frac{1}{2}(\Tr U)^2\Tr U^{n-2}
\nn\\&& + 
      \frac{1}{2}\Tr U^2 \Tr U^{n-2}
  + \Tr U \Tr U^{n-1}
\qquad \forall U\in \rm{SU}(8)
\eea

\bibliography{thesis-paper3}
\end{document}